\documentclass[oldversion]{aa}
\usepackage{graphicx}
\usepackage{natbib}
\usepackage{hyperref}
\usepackage{longtable}
\usepackage{lscape}
\begin{document}

   \title{Dynamical masses of M-dwarf binaries in young moving groups: II -- Toward empirical mass-luminosity isochrones
\thanks{Based on observations collected at the European Southern Observatory, Chile (Programmes 090.C-0819, 60.A-9386, 098.C-0262, and 099.C-0265).}
}

  % \subtitle{Young low-mass binaries}

   \author{Markus Janson\inst{1} \and
          Stephen Durkan\inst{1,2} \and
          Micka{\"e}l Bonnefoy\inst{3} \and
          Laetitia Rodet\inst{3} \and
          Rainer K{\"o}hler\inst{4,5}
          Sylvestre Lacour\inst{6} \and
          Wolfgang Brandner\inst{7} \and
          Thomas Henning\inst{7} \and
          Julien Girard\inst{8}
          }

   \institute{Department of Astronomy, Stockholm University, Stockholm, Sweden\\
              \email{markus.janson@astro.su.se}
         \and
            Astrophysics Research Center, Queens University Belfast, Belfast, Northern Ireland, UK
        \and
            Univ. Grenoble Alpes, IPAG, Grenoble, France
        \and
            Institut f{\"u}r Astro- und Teilchenphysik, Universit{\"a}t Innsbruck, Innsbruck, Austria
        \and
            Department of Astrophysics, University of Vienna, Vienna, Austria
        \and
            LESIA, Observatoire de Paris, Meudon, France
        \and
            Max Planck Institute for Astronomy, Heidelberg, Germany
        \and
            Space Telescope Science Institute, Baltimore, Maryland, USA
             }

   \date{Received ---; accepted ---}

   \abstract{Low-mass stars exhibit substantial pre-main sequence evolution during the first $\sim$100 Myrs of their lives. Thus, young M-type stars are prime targets for isochronal dating, especially in young moving groups (YMGs), which contain large amounts of stars in this mass and age range. If the mass and luminosity of a star can both be directly determined, this allows for a particularly robust isochronal analysis. This motivates in-depth studies of low-mass binaries with spatially resolvable orbits, where dynamical masses can be derived. Here we present the results of an observing campaign dedicated to orbital monitoring of AB Dor Ba/Bb, which is a close M-dwarf pair within the quadruple AB Dor system. We have acquired eight astrometric epochs with the SPHERE/ZIMPOL and NACO instruments, which we combine with literature data to improve the robustness and precision for the orbital characterization of the pair. We find a system mass $0.66^{+0.12}_{-0.12}$~$M_{\rm sun}$ and bolometric luminosities in $\log{L/L_{\rm sun}}$ of $-2.02\pm0.02$ and $-2.11 \pm 0.02$ for AB Dor Ba and Bb, respectively. These measurements are combined with other  YMG pairs in the literature to start building a framework of empirical isochrones in mass--luminosity space. This can be used to calibrate theoretical isochrones and to provide a model-free basis for assessing relative stellar ages. We note a tentative emerging trend where the youngest moving group members are largely consistent with theoretical expectations, while stars in older associations such as the AB Dor moving group appear to be systematically underluminous relative to isochronal expectations.}


\titlerunning{Orbit of AB Dor Ba/Bb}
\authorrunning{M. Janson et al.}

   \maketitle
%
%________________________________________________________________

\section{Introduction}
\label{s:intro}

Stellar systems that are both young and nearby are of importance for a range of present-day scientific topics, not least for the purpose of direct imaging of exoplanets \citep[e.g.][]{marois2008,macintosh2015,chauvin2017} and disks \citep[e.g.][]{schneider2009,thalmann2013,boccaletti2015}. This has led to an increased interest in young moving groups (YMGs), which are associations of stars that are unbound but clustered in phase space, and thus are expected to have originated from a mutual birth cluster \citep[e.g.][]{torres2008}. One such group that is particularly close, and thus particularly useful for many purposes, is the AB Dor Moving Group (ABMG) \citep[e.g.][]{zuckerman2004}. While ABMG is clearly older than 5--20~Myr, which is the approximate age of the youngest YMGs such as the TW Hya or $\beta$~Pic associations \citep[e.g.][]{bell2015}, its specific age has remained uncertain, with different studies suggesting age ranges from a lower limit of 30~Myr \citep{close2005} all the way to an upper limit of 200~Myr \citep{bell2015}.

The defining member of the ABMG, AB~Dor itself, is a complex and intriguing system. The primary AB Dor A is a K-type star, which has long been known to share a common proper motion with the M-type secondary AB Dor B \citep{rossiter1955} at a separation of $\sim$10$^{\prime \prime}$. However, more recently it has been discovered that A and B can each be resolved into tight stellar pairs. AB Dor C is a $\sim$90$M_{\rm jup}$ star near the hydrogen burning limit in a 11.75-year orbit around AB Dor A \citep{guirado1997,close2005,azulay2017b}. AB Dor B is in fact a nearly equal-mass stellar pair \citep{janson2007} designated as AB Dor Ba and Bb. The Ba/Bb pair has been the subject of particular attention in several studies, due to its particular properties. Both stars are M5--M6 type stars, which means that unlike earlier-type stars, they still reside in the pre-main sequence (PMS) phase at the age of the ABMG. Furthermore, orbital monitoring of the system \citep[e.g.][hereafter W14 and A15, respectively]{wolter2014,azulay2015} has shown that the orbital period is only $\sim$1 year, which benefits the determination of precise stellar masses and ages. \citet{azulay2015} estimate masses of 0.28$\pm$0.05~$M_{\rm sun}$ and 0.25$\pm$0.05~$M_{\rm sun}$ for the Ba and Bb components, respectively. This implies a total mass 23\% lower than the 0.69~$M_{\rm sun}$ derived by W14, although the two estimates are consistent within the errors. Given that the observations of both previous studies largely cluster around the apoapsis of the orbit where the binary spends most of its time, these parameters can be further constrained by targeting previously unobserved orbital phases. The A15 study is based on radio interferometry using a quasar as phase reference, and could therefore also produce an absolute parallax of 66.4$\pm$0.5~mas, corresponding to a distance of 15.06$\pm$0.11~pc. This is consistent within the errors with the Hipparcos parallax for AB Dor A of 66.92$\pm$0.54~mas \citep{perryman1997} and marginally consistent with the Gaia DR2 parallax of AB Dor B of 67.03$\pm$0.09~mas \citep{brown2018}, which has a better formal precision. However, here we adopt the A15 parallax for AB for Ba/Bb because the Gaia measurement may conceivably be affected by the Ba/Bb orbit (and likewise, the AB Dor A component could be affected by the A/C orbit).

 M dwarfs remain for a long time in the pre-main sequence phase, and  dynamical masses allow for direct comparison between observational data and isochronal models, which make M-dwarf binaries in YMGs an important sample for astrophysical calibrations \citep{janson2017}. In this context, AB Dor Ba and Bb are of particularly high priority, given their short orbital period, which makes them promising for calibrating theoretical models of young stars, as well as for potentially constraining the age of the system if well matching isochrones are found. This could in turn have implications for the entire ABMG. Independent of any model uncertainties, dynamic and photometric/spectroscopic data of the binary can also be used to define an empirical isochrone, which through comparison with other binaries in ABMG or other YMGs can provide precise information about the relative ages of different YMGs, or the age spread within individual YMGs. Thus, we have performed a dedicated study of the astrometry for this system with the goal of further constraining the orbital properties by adding data over a larger orbital phase coverage and with higher precision than has been previously available with near-infrared imaging. Here, we present the results of this study.

The article is organized as follows. In section \ref{s:obs}, we outline the observations and reduction of the data included in this study. This data is used for astrometric extraction, which is described in Sect. \ref{s:astro}, and which in turn is used for orbit fitting to constrain the AB Dor Ba/Bb orbital parameters, as discussed in Sect. \ref{s:orbit}. Since the main aim of the study is isochronal analysis in mass-luminosity space, we also need to derive bolometric luminosities, which is the topic of Sect. \ref{s:lbol}. The isochronal analysis itself is  described in Sect. \ref{s:iso}, where we relate our result to other results of YMG binaries in the literature in order to build a framework for empirical isochrones in mass versus luminosity. Finally, the results are summarized in Sect. \ref{s:summary}.

\section{Observations and data reduction}
\label{s:obs}

This study considers both archival and newly acquired data for the purpose of constraining the AB Dor Ba/Bb orbit. The archival data includes NACO\footnote{NAOS-CONICA \citep{lenzen2003,rousset2003}.} images from various programmes compiled by W14, and VLBI radio interferometric data presented in A15. Our new data includes Sparse Aperture Masking (SAM) observations with NACO, offering a higher angular resolution than archival NACO images, and a sequence of images from the ZIMPOL\footnote{Zurich Imaging Polarimeter \citep{thalmann2008}.} arm of the SPHERE \footnote{Spectro-Polarimetric High-contrast Exoplanet Research \citep{beuzit2008}.} instrument. ZIMPOL enables a rather high Strehl ratio (often in excess of 50\%) even for wavelengths as short as R-band, and thus provides unparalleled angular resolution for full aperture imaging. 

The ZIMPOL observations are the most recent (from 2016--2017, apart from a test epoch taken during SVT\footnote{Science Verification Time}), and their scheduling was informed by the previously acquired astrometry for the system. Since the period was known to be very close to one year, the ZIMPOL scheduling was spread out over one year, with denser sampling during phases of the orbit that were not previously well covered. At $-65$ deg latitude, AB Dor B is effectively circumpolar, but from approximately late April to mid July, the airmass for AB Dor B as seen from Paranal is too high ($>$1.8) to acquire a sufficient image quality during night time. Thus, the ZIMPOL observations are restricted to a $\sim$9-month window during the year. Since two of the requested observations could not be executed by the observatory within their required time windows, our total coverage was in practice limited to 6 months. Still, the observations cover a significant fraction of the orbital phase that had previously not been probed. All observations are summarized in Table \ref{t:log}. In principle, another NACO epoch from 2008.65 exists, as presented in W14. However, that data point is a special case where only a marginal PSF extension was seen in the NACO image. The fitting performed in W14 for this epoch also provided a flux ratio that was inconsistent with their other values. They therefore concluded that the derived separation of 19 mas should merely be regarded as an upper limit to the separation. Here, we note that 19 mas corresponds to only $\sim$0.3~$\lambda / D$ for the VLT in $K_{\rm s}$ band, so the measurement would be challenging even for a perfectly diffraction-limited PSF. With the imperfect Strehl ratio and stability of the NACO PSF, we estimate that there is a risk that the measured extension could be substantially affected by PSF imperfections. We thus opt to omit the 2008.65 data point in our analysis.

% Table: Date, Instrument/Mode, Band, Reference
\begin{table}[htb]
\caption{Archival and new observations of AB Dor Ba/Bb}
\label{t:log}
\centering
\begin{tabular}{lll}
\hline
\hline
Date & Facility & Reference \\
\hline
2004-02-03      & NACO   & W14 \\
2005-01-07      & NACO   & W14 \\
2005-11-28      & NACO   & W14 \\
2007-11-11      & VLBI   & A15 \\
2008-11-08      & NACO   & W14 \\
2008-12-19      & NACO   & W14 \\
2009-01-01      & NACO   & W14 \\
2009-02-17      & NACO   & W14 \\
2010-10-25      & VLBI   & A15 \\
2012-11-26      & NACO-SAM       & This paper \\
2012-12-20      & NACO-SAM       & This paper \\
2013-01-28      & NACO-SAM       & This paper \\
2013-08-16      & VLBI   & A15 \\
2014-12-11      & ZIMPOL         & This paper \\
2016-10-08      & ZIMPOL         & This paper \\
2017-01-13      & ZIMPOL         & This paper \\
2017-02-16      & ZIMPOL         & This paper \\
2017-03-06      & ZIMPOL         & This paper \\
\hline
\end{tabular}
%\tablenotetext{a}{No notes at the moment.}
\end{table}

All the NACO-SAM observations were obtained with the same setting: the $K_s$ filter, the S27 camera (27 mas pixel scale), and the seven-hole mask \citep{tuthill2010}. Observations of AB Dor Ba/Bb were interspersed with several observations of three different calibrators: HD\,35936, HD\,37364, and HD\,271187. To speed up the process, we used the star-hopping technique where the calibrator is reached by means of a simple offset of the telescope, without re-optimization of the AO system. Each individual exposure is composed of eight individual data cubes of 151 0.3-second frames, dithered over the four quadrants of the detector. Each detector quadrant is sky subtracted using the six data cubes where no source is present.

With the exception of the SVT epoch, all ZIMPOL data were taken with the same observational settings. We used the $N_R$ filter with a central wavelength of 645.9~nm and bandwidth of 56.7~nm since this allows for an efficient distribution of light between the science camera and the wavefront sensor using the dichroic beamsplitter. Twenty-five frames were acquired per epoch, with two readouts per frame and 55 seconds per readout. We also had originally intended  for the SVT run to be executed with the $N_R$ filter, but during execution the $H_{\alpha}$-NB filter was mistakenly used instead, leading to much lower fluxes than intended in the images. However, the data were still of sufficient quality to derive reasonably precise astrometry for this epoch. The SVT observations encompassed five frames with five readouts per frame and 50 seconds per readout. All observations were acquired in service mode, with the constraint that the seeing had to remain below 0.8$^{\prime \prime}$. This was mostly fulfilled, except in the January epoch where the seeing fluctuated considerably during the run.  However, since some of the data were of sufficient quality, this epoch could still be used for precise astrometry. All data were acquired in field stabilized mode.

\begin{figure*}[htb]
\centering
\includegraphics[width=18cm]{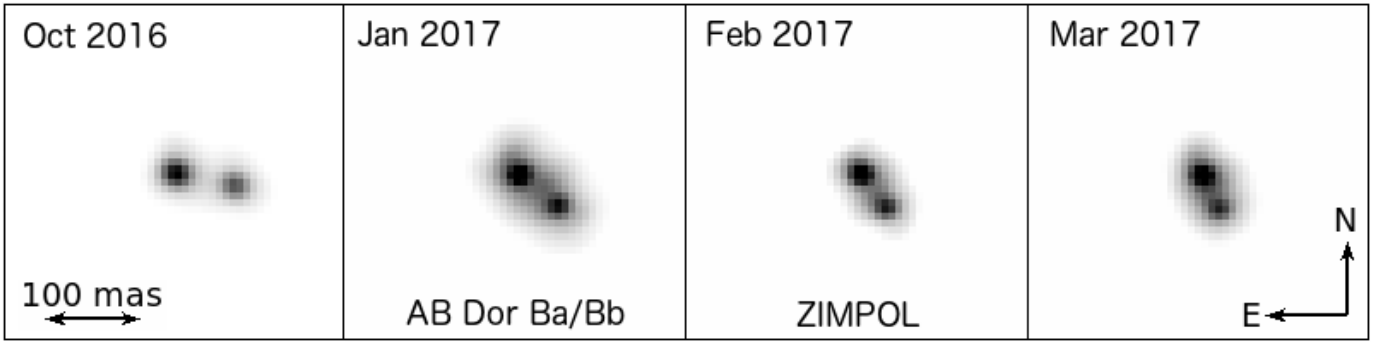}
\caption{Series of images of the AB Dor Ba/Bb system from ZIMPOL taken from Oct 2016 to Mar 2017. While the two stellar components were only partially resolvable with previous generations of AO systems, the high Strehl ratio and high spatial resolution of ZIMPOL allows  the components to be clearly resolved.}
\label{f:img}
\end{figure*}

ZIMPOL pixels have a tentative pixel scale of 3.601$\pm$0.005 mas/pixel \citep{schmid2017}, but only every other row is read out on the sky in each individual frame. In order to acquire an equal and uniform sampling in the $x$ and $y$ directions without interpolations, we simply downsampled to an effective pixel scale of 7.202 mas/pixel in both directions. We performed bias correction on individual rows and in individual image quadrants based on the reference values given in the edges of the frames, and corrected for flat-field effects using a lamp flat. An example ZIMPOL frame is shown in Fig. \ref{f:img}

During the ZIMPOL programme, we also observed GJ~3323 in one epoch to serve as a point spread function (PSF) reference star. However, GJ 3323 was spatially extended in the images, possibly as a result of previously unresolved binarity. As a result, and as originally intended as another approach for this purpose, we instead used archival ZIMPOL PSF reference stars, which are regularly observed as part of the standard ZIMPOL calibration package. We used three $N_R$ `flux standard' images from 29 October 2015, 30 April 2016, and 7 May 2016 for this purpose. The PSF standards were reduced in the same way as for the science data as described above. 

%Figure with ZIMPOL image. 

\section{Astrometric extraction}
\label{s:astro}

The astrometric extraction from the SAM data is performed using the closure phases of the interferometric pattern. Complex visibilities are retrieved using the SAMP pipeline \citep{lacour2011}: bad pixels are flagged, the sky is subtracted, frames are centred, and the fringe pattern is fitted using a theoretical diffraction pattern of the seven-hole mask. The 21 complex visibilities of AB Dor Ba/Bb are then calibrated using the 21 complex visibilities of the calibrators. Calibrators are cross-calibrated to check that they are indeed single stars. Finally, the astrometric fit is done after computation of the closure phase using a two-component (binary) model. Errors are normalized to achieve a residual chi-square of one. True north correction and pixel scale were calculated in the same way as in \citet{chauvin2012} on the calibration target $\theta_1$~Ori~C, which was observed on 28 January 2013, yielding a pixel scale of 27.03$\pm$0.18 mas/pixel and a true north correction of 0.43$\pm$0.30 deg.

For  the ZIMPOL data, we used an iterative PSF fitting scheme to account for the partially overlapping PSFs of the AB Dor Ba and Bb components. This is the same procedure as previously developed for the AstraLux M-dwarf multiplicity survey \citep{janson2012,janson2014a}. In brief, a rough initial estimation is made for the locations and brightnesses of the two stars. A model system is then built using two copies of a PSF reference star. The positions and brightnesses of the two components are then iteratively varied until a minimum residual solution (relative to the target image) is found. For AB Dor Ba/Bb, we run this procedure for all individual readouts for each epoch in order to check the scatter among individual images. Furthermore, the procedure is run with each of the three different PSF references in order to evaluate the error resulting from PSF matching imperfections. Due to the relative faintness of AB Dor B at visible wavelengths (R$\sim$11 mag), the adaptive optics correction is sometimes unstable, so in order to sort out low-quality frames, we only perform fits to those frames in which the brightness within a 10-pixel radius aperture from the photocentre of the binary exceeds 50\% of the corresponding value in the brightest frame.

A PSF that is affected by the atmosphere is effectively a superposition of two components: A diffraction-limited core, and a seeing-limited halo. Diffraction is independent of ambient conditions while seeing is not, so the shape of the halo will vary much more strongly with time than the shape of the core. Thus, to benefit the matching of a PSF reference star to a target PSF, it is often useful to filter out the halo with high-pass filtering (e.g. unsharp masking) so that the fit is made almost exclusively on the core component. Thus, unsharp masking using a broad Gaussian kernel is a standard procedure for Lucky Imaging data \citep[e.g.][]{bergfors2010}. For ZIMPOL, the Strehl ratio is rather high, such that a large fraction of the light is already concentrated into the core, making the benefit from usage of unsharp masking less immediately obvious. Thus, we performed a test with two identical fitting procedures for all stars, with the exception that one underwent unsharp masking using a Gaussian kernel of full width at half maximum (FWHM) of 15 pixels, while the other did not. The results were fully consistent within the error bars with no signs of any systematic differences. However, the unsharp masking data had slightly smaller errors (as expected), hence we used them for the analysis in the following. 

\begin{figure*}[htb]
\sidecaption
\includegraphics[width=12cm]{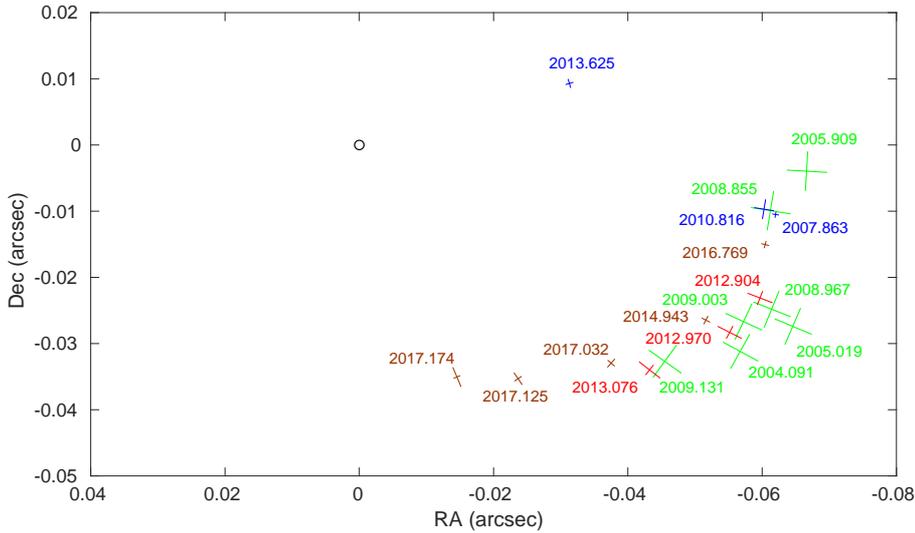}
\caption{Astrometric data of AB Dor Bb relative to AB Dor Ba, collected with ZIMPOL (brown) and NACO-SAM (red), along with literature data from W14 (green) and A15 (blue), labeled with the epochs of each observation. The black circle indicates the fixed location of AB Dor Ba.}
\label{f:astro}
\end{figure*}

The PSF fitting procedure outputs differential coordinates $\delta x$ and $\delta y$, and the differential brightness $\delta m$ between AB Dor Ba and AB Dor Bb, with uncertainties represented by the standard errors from frame-to-frame scatter and PSF-to-PSF reference scatter added in quadrature. These results are in pixels, so to convert into sky coordinates we apply two times the tentative pixel scale of 3.601$\pm$0.005 mas/pixel (since our data are rebinned by a factor of 2), and the true north correction of -2.0$\pm$0.5 deg \citep{schmid2017}. We note that the true north correction here refers to the angle of the vertical axis relative to north; for example, an angle of  90 deg relative to the vertical axis corresponds to an angle of 88 deg relative to north. The derived astrometric quantities are summarized in Table \ref{t:astro}. We show them along with the literature data in Fig. \ref{f:astro}. The various classes of data are broadly consistent, although some interesting trends can be seen, for example  the projected separation in the literature NACO data appear to have a larger average projected separation than the NACO-SAM data, which in turn seem to have a larger average separation than the ZIMPOL data. This is probably due to systematic effects in the astrometric calibration, although we also discuss alternative interpretations in Sect. \ref{s:nonkeplerian}.

%Figure with astrometric data from different instruments in different colors.

% Table: Date, sep, PA, band
\begin{table}[htb]
\caption{Astrometric results.}
\label{t:astro}
\centering
\begin{tabular}{llll}
\hline
\hline
Date & Sep. (mas) & PA (deg) & Instrument \\
\hline
2012-11-26      & 64$\pm$2 & 248.8$\pm$1.0 & NACO-SAM \\
2012-12-20      & 62$\pm$2 & 242.8$\pm$1.0 & NACO-SAM \\
2013-01-28      & 55$\pm$2 & 231.8$\pm$1.0 & NACO-SAM \\
2014-12-11      & 58.0$\pm$0.6 & 242.86$\pm$0.69 & ZIMPOL \\
2016-10-08      & 62.3$\pm$0.6 & 256.02$\pm$0.51 & ZIMPOL \\
2017-01-13      & 49.9$\pm$0.7 & 228.69$\pm$0.97 & ZIMPOL \\
2017-02-16      & 42.5$\pm$1.1 & 213.81$\pm$0.88 & ZIMPOL \\
2017-03-06      & 37.9$\pm$1.6 & 202.53$\pm$0.82 & ZIMPOL \\
\hline
\end{tabular}
%\tablenotetext{a}{No notes at the moment.}
\end{table}

\section{Orbital constraints}
\label{s:orbit}

We  performed orbital fitting to the relative Ba/Bb astrometry for the ZIMPOL and NACO-SAM data in combination with the literature data using on one hand the procedure developed and described in \citet{kohler2008,kohler2012} based on the Levenberg-Marquardt algorithm \citep{press1992}, and on the other hand a routine described in \citet{chauvin2012} based on a Markov chain Monte Carlo (MCMC) Bayesian analysis technique \citep{ford2005,ford2006}. The two routines give consistent best-fit values, but the errors are larger in the MCMC fit. We consider the MCMC procedure more conservative in this regard since it accounts for a wide range of correlations among the parameters, and use its output for all the  analyses in the following. 

\begin{figure*}[htb]
\sidecaption
\includegraphics[width=12cm]{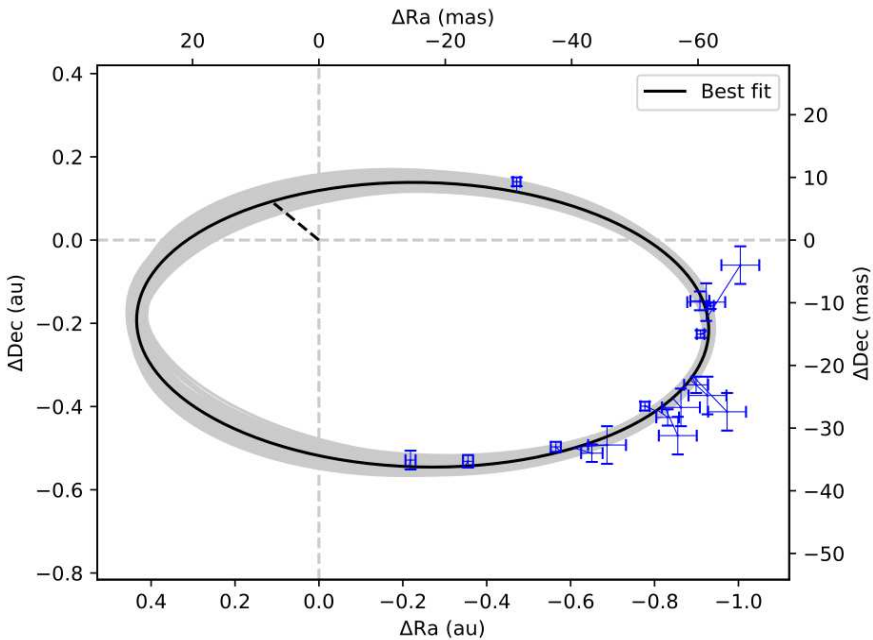}
\caption{Resulting orbit of AB Dor Ba/Bb from our orbital fitting.}
\label{f:orbit}
\end{figure*}

\begin{figure*}[htb]
\centering
\includegraphics[width=16cm]{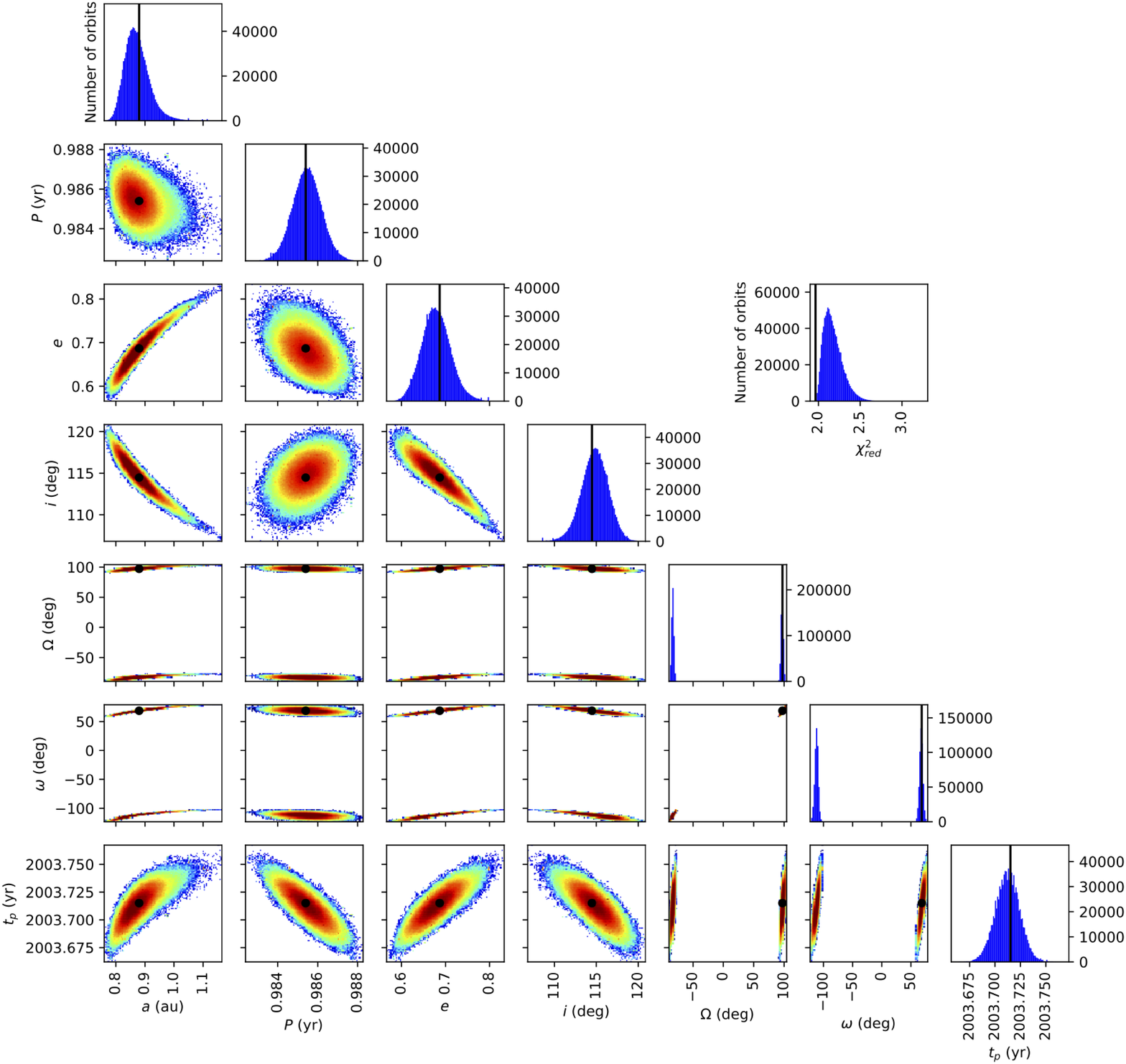}
\caption{Posterior distributions of our fitted parameters, along with their mutual correlations.}
\label{f:posteriors}
\end{figure*}

The resulting fit is shown in Fig. \ref{f:orbit}, the posteriors on the orbital parameters and their mutual correlations are shown in Fig. \ref{f:posteriors}, and the best-fit parameters are listed in Table \ref{t:stdorbit}. Most of the parameters are consistent with those reported in A15 within the mutual error bars but with considerably smaller errors in our fit, as expected considering the new and precise astrometric data points provided in Sect. \ref{s:astro}. The semi-major axis is also consistent with that in A15; however, in this case, the error is  larger in our fit. The reason for this is most likely our use of MCMC for the fitting, which is more robust to parameter correlations than the Levenberg-Marquardt algorithm used in A15. Another reason may be that A15 use the 2008.65 NACO data point from W14 in their analysis, which we omit as discussed in Sect. \ref{s:obs}. The larger error in semi-major axis also leads  to a larger error in mass than in A15. Our derived mass is consistent with both W14 and A15 within error bars. The total system mass is derived as 0.66$\pm$0.12~$M_{\rm sun}$. 

% For all the astrometric values, we follow the A15 procedure, not adopting the more recent 

%Figure with best orbital fit.

%Figure with best orbital fit.

Thanks to the long baseline relative to the orbit and the precise measurements, the period can now be constrained to significantly better precision than 1 day, and has a fractional error of $\sim$0.1\%. Of the three components that play a  role in deriving a total system mass (angular semi-major axis $\alpha$, distance $d$, and period $P$), $P$ has the weakest exponent and the smallest fractional error, so it is now  constrained closely enough to have a negligible impact on the system mass uncertainty. Of the two remaining parameters, which in combination yield the physical semi-major $a$, $\alpha$ has a fractional error of $\sim$6\% and the parallax-based distance has 0.7\%. Hence, $\alpha$ is the limiting factor in the mass determination, and any future attempts to further constrain the orbit are best focused on placing tighter constraints on this parameter, particularly by collecting data even closer to periastron.

\begin{table}
\caption{Parameters of the best orbital solution.}
\label{t:stdorbit}
%% stretch a bit so that sub- and superscripts don't overlap
\renewcommand{\arraystretch}{1.3}
\begin{center}
\begin{tabular}{lr@{}l}
\noalign{\vskip1pt\hrule\vskip1pt}
Orbital Element                         & \multicolumn{2}{c}{Value} \\
\noalign{\vskip1pt\hrule\vskip1pt}
Date of periastron $T_0$ (yr)                   & $2003.713$ & $\pm 0.015$\\
                                                & (2003 Aug 16)\span\\
Period $P$ (yr)                         & $0.9856$ & $\pm 0.0009$\\
Period $P$ (d)                          & $359.98$ & $\pm 0.33$\\
Angular semi-major axis $\alpha$ ($^{\prime \prime}$)                       & $0.0571$ & $\pm 0.0034$\\
Semi-major axis $a$ (AU)                        & $0.86$ & $\pm 0.05$\\
Eccentricity $e$                                & $0.669$ & $\pm0.039$\\
Argument of periastron $\omega$ ($^\circ$)      & $68.0$ & $\pm 3.5$\\
P.A. of ascending node $\Omega$ ($^\circ$)      & $96.8$ & $\pm 2.2$\\
Inclination $i$ ($^\circ$)                      & $114.9$ & $\pm 1.9$\\
System mass $M_{\rm S}$ ($M_{\rm sun}$)                   & $0.66$ & $\,^{+0.12}_{-0.12}$\\
\noalign{\vskip1pt\hrule\vskip1pt}
\end{tabular}
\end{center}
\end{table}

\subsection{Non-Keplerian alternatives}
\label{s:nonkeplerian}

In a previous section we noted that there are indications for small but consistent offsets between different data sets, which we interpreted as residual systematic errors. However, it is also relevant to evaluate whether there could be an actual physical effect at play in the observed system to cause these offsets. In particular, the orbit fitting described above assumes that the AB Dor Ba/Bb system consists of two bodies in a closed Keplerian orbit, but if any third body in the system has a sufficiently large gravitational impact on the pair, this assumption breaks down. In this context, it is especially relevant to note that around apastron, the W14 data points (2004--2009) have a larger mean separation than the NACO-SAM points (2012--2013), which in turn have a higher mean separation than the ZIMPOL points (2014--2017). This could in principle be interpreted as continuous orbital evolution over $\sim$1~yr timescales.

To evaluate this scenario, we first turn our attention to the known additional component of the system,   the AB Dor A/C pair, which is dominated in mass by the A component. At a present-day separation of at least 150~AU from AB Dor Ba/Bb, it is clear that AB Dor A cannot impose dynamical changes on a 1~yr timescale. Even at large separations, it can impose Kozai-Lidov \citep{kozai1962,lidov1962} oscillations on the Ba/Bb pair, if it is sufficiently inclined to the Ba/Bb orbital plane. However, the timescale for Kozai-Lidov oscillations is proportional to $P_{A/B}^2 / P_{Ba/Bb}$ \citep[e.g.][]{antognini2015}, where $P_{Ba/Bb}$ is close to 1~yr as determined in our orbital fitting, and $P_{A/B}^2$ is of order $10^6$ yr$^2$ or more at 150~AU projected separation, so it is highly unlikely that it could have any impact on the timescales we are considering here.

Next we consider additional components within the AB Dor Ba/Bb system,   i.e. unresolved companions around either star. Such companions could cause an impact on the observed Ba/Bb orbit either through three-body dynamical interactions, forcing the orbit of the visible components to evolve over time, or through simply shifting the photocentre of its host star during its orbit. The latter can again be excluded on the basis of the timescale. The orbital period of such an unseen companion would have to be much shorter than the 360~d orbit of Ba/Bb, so we would expect its astrometric effects to manifest itself as  scatter on short timescales, such as within the ZIMPOL 2016--2017 arc, rather than a consistent offset between the ZIMPOL and NACO-SAM arcs. The lack of scatter within these arcs may also be a problem for the three-body dynamics hypothesis, although in this case the solution may be fine-tuned by assigning a small mass to the unseen companion such that it has a negligible impact on the photocentre on the component it orbits, but still has close enough encounters with the other component that it inflicts a large dynamical disturbance. However, such a configuration should be extremely short-lived since the less massive perturber should be rapidly ejected from the system, so it is an unlikely scenario for that reason. Taking all these arguments into account, we thus argue that non-Keplerian solutions are very improbable, and that the small remaining offsets seen in the data are instrumental effects rather than any physical mechanism at play in the AB Dor system.

\section{Luminosity constraints}
\label{s:lbol}

By relating the dynamical mass of our stellar components to a quantity that is expected to evolve with time, we can perform an isochronal analysis to either produce a model-dependent age estimate, or conversely, to test the accuracy of the isochronal models with an independent age estimate. The bolometric luminosity $L_{\rm bol}$ is particularly useful in this context. For the purpose of estimating $L_{\rm bol}$, we first acquire unresolved photometry from the literature at a range of wavelengths. The \textit{Gaia} $G$ filter \citep{prusti2016} is used for the visible wavelength range, since it covers a wide band with a small photometric error. For the near-infrared regime we use 2MASS \citep{skrutskie2006} $JHK$ bands, and for the mid-infrared we use the WISE \citep{wright2010} $W1$--$W4$ bands. Furthermore, we measure differential photometry in the $N_R$ and $K$ bands in our data. We then construct a grid of theoretical binaries based on the BT-SETTL models \citep{allard2014}. Each artificial component is given a mass in the range of 0.1--1~$M_{\rm sun}$ and an age in the range of 50--150~Myr, and every possible pairing of components is used to produce a prediction of the total brightness and differential brightness in the bands listed above. If a given artificial pair matches the observations within uncertainties, then the predicted $L_{\rm bol}$ values of the individual components are considered to be part of the set of bolometric luminosities that accurately represent the real values. The mean and scatter of these sets then represent the adopted $L_{\rm bol}$ for the two components. All the photometric values and derived luminosities are summarized in Table \ref{t:mag}. The advantage of this method is that it uses photometric information across a wide range of wavelengths, sampling a large fraction of the stars' total bolometric outputs. In this way, it effectively averages over all different atmospheric and physical properties that could potentially represent the system, making the result quite insensitive to specifics among these effects (e.g. the degree of atmospheric absorption within a given photometric band). 

% Table: Brightness and luminosity
\begin{table}[htb]
\caption{Photometry and luminosity of AB Dor Ba/Bb}
\label{t:mag}
\centering
\begin{tabular}{lll}
\hline
\hline
Quantity &Value & Reference \\
\hline
Combined $G$ (mag)      & 11.159$\pm$0.002       & \textit{Gaia} \\
Combined $J$ (mag)      & 8.171$\pm$0.018        & 2MASS \\
Combined $H$ (mag)      & 7.659$\pm$0.042        & 2MASS \\
Combined $K$ (mag)      & 7.341$\pm$0.031        & 2MASS \\
Combined $W1$ (mag)     & 7.038$\pm$0.191        & WISE \\
Combined $W2$ (mag)     & 6.897$\pm$0.103        & WISE \\
Combined $W3$ (mag)     & 6.852$\pm$0.043        & WISE \\
Combined $W4$ (mag)     & 6.744$\pm$0.094        & WISE \\
$\delta N_R$ (mag)      & 0.33$\pm$0.05  & This paper \\
$\delta K$ (mag)                & 0.24$\pm$0.14  & This paper \\
$\log{L_{\rm bol, Ba} / L_{\rm sun}}$   & -2.02$\pm$0.02         & This paper \\
$\log{L_{\rm bol, Bb} / L_{\rm sun}}$   & -2.11$\pm$0.02         & This paper \\
\hline
\end{tabular}
%\tablenotetext{a}{No notes at the moment.}
\end{table}

One of the scientific aims of this study is to compare the results for AB Dor Ba/Bb with other low-mass YMG binaries in the literature;  therefore, we have collected the relevant isochronal quantities for all low-mass components in YMG binaries with good individual dynamical mass constraints that we are aware of. Masses, and in some cases luminosities, come from \citet{close2007} and \citet{azulay2017b} for AB Dor AC; \citet{montet2015} for GJ 3305 AB; \citet{kohler2013} and \citet{kohler2018} for TWA 5 AB; \citet{nielsen2016} for V343 Nor B; \citet{calissendorff2017} for 2MASS J10364483+1521394 BC (hereafter J1036); \citet{azulay2017a} for HD 160934 AC; and \citet{rodet2018} for GJ 2060 AB. For V343 Nor B and J1036 BC, no $L_{\rm bol}$ was available, and for HD160934 AC, the value quoted in \citet{hormuth2007} is based on the distance from the original Hipparcos reduction, which is inconsistent with newer values such as the VLBI parallax in \citet{azulay2017a}; for these cases we calculated an $L_{\rm bol}$ in the same way as described for AB Dor Ba/Bb above, with small variations depending on which unresolved and resolved photometric data points were available for each target.  We find $\log{L_{\rm bol} / L_{\rm sun}}$ of -1.56$\pm$0.07 for V 343 Nor B, -0.97$\pm$0.02 for HD 160934 A, -1.40$\pm$0.03 for HD 160934 C, and -2.49$\pm$0.02 for both J1036 B and C. 

\section{Isochronal analysis}
\label{s:iso}

\begin{figure*}[htb]
\sidecaption
\includegraphics[width=12cm]{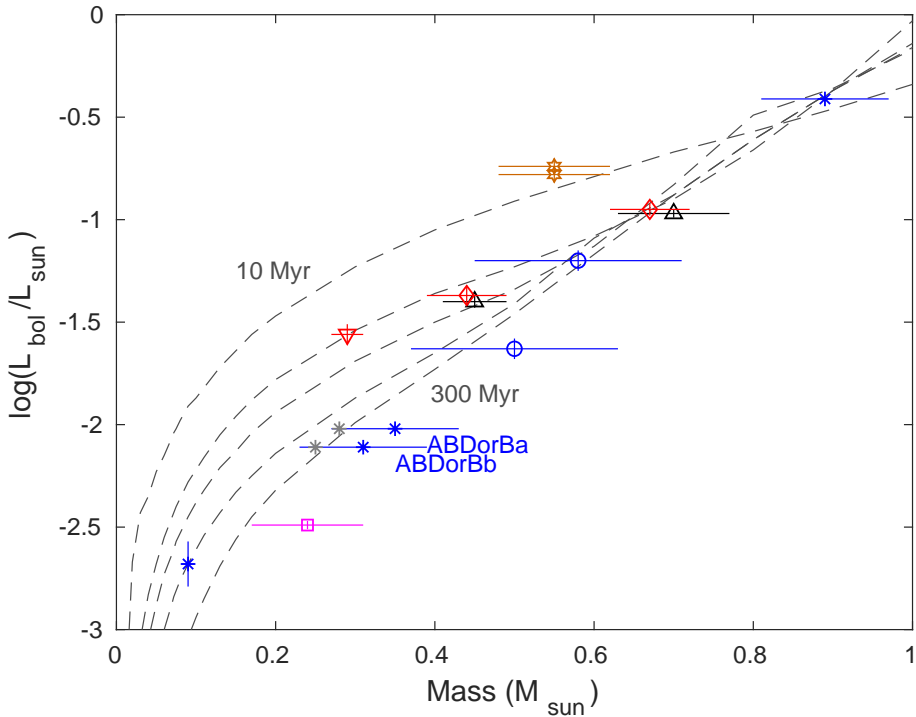}
\caption{Isochronal analysis of AB Dor Ba/Bb and other low-mass YMG binaries with known orbits. Brown symbols denote TWA members (TWA 5 AB), red are BPMG members (diamonds: GJ 3305 AB, triangle: V343 Nor B), blue are ABMG members (asterisks: AB Dor system, circles: GJ 2060 AB), and magenta are UMaMG members (J1036 BC; the two stars appear as one point since they have equal derived properties). Black symbols denote unclear YMG membership (HD 160934 AC). For AB Dor Ba/Bb, the original A15 masses are plotted as grey asterisks, while the masses scaled for the total mass derived in this work are plotted in blue. Also plotted are BCAH15 isochrones going from 10 Myr (top) through 30, 50, 100, and 300 Myr (bottom). The last (300 Myr) is already close to the main sequence at these masses, and thus largely indistinguishable from older ages.}
\label{f:iso}
\end{figure*}

The masses and luminosities of AB Dor Ba and Bb are plotted in Fig. \ref{f:iso}, along with the corresponding measurements from relevant binaries in the literature. Individual masses for the two components are presented in A15, but this predicts a lower total mass than we find in our fitting of the relative orbit. Thus, we show two separate individual masses for each component in Fig. \ref{f:iso}, where one corresponds to the exact masses in A15 and the other  to the same mass ratio as A15 but scaled to account for the difference in total mass. As discussed previously, the total masses are consistent within the errors. Also plotted are the corresponding measurements from relevant binaries in the literature. By now, a sufficiently large sample of these stars is emerging to  start building a framework of empirical isochrones. In general,  stars of a common age are expected to fall approximately along a line in a mass--luminosity diagram, and thus constitute an empirical isochrone. This isochrone could be fit with theoretical isochrones, which can provide an age estimate for the stars based on which isochrone fits the best or, conversely, can be used to calibrate the theoretical isochrones, for example by distinguishing which out of a set of different theories provides the best fit or by identifying parameter ranges in which the models fit comparatively poorly. However, empirical isochrones of different populations of stars can also be useful completely irrespective of theoretical models since it provides a framework for determining model-free relative ages. For example, the isochrones of two kinematically distinct YMGs can be used to evaluate which one is younger, and the scatter around an isochrone within a YMG can provide clues about the degree of age spread within the group. Among the sample we have included in Fig. \ref{f:iso}, TWA 5 is associated with the TW Hya (TWA) YMG with an age of $\sim$5--10~Myr; GJ 3305 and V343 Nor B are associated with the $\beta$ Pic moving group (BPGM) with an age of $\sim$20--30~Myr; AB Dor and GJ 2060 are associated with the AB Dor moving group (ABMG) with an age of $\sim$50--150~Myr; and J1036 has been identified as an UMa moving group (UMaMG) member with an age in the range of 300--500~Myr \citep[e.g.][]{brandt2015,jones2015}. HD 160934 has been identified as an ABMG member \citep{lopez2006}; this classification has varied substantially in the literature \citep[see e.g.][]{azulay2017a} and across different versions of the BANYAN tool \citep[e.g.][]{gagne2014,gagne2018},  and we thus consider it  an unclear case. The M-dwarf pair TWA~22 also has a tight system mass constraint \citep{bonnefoy2009,rodet2018}, but we do not include it in this analysis since the mass ratio is not yet determined.

%Empirical and theoretical isochrones.

Several interesting trends emerge in Fig. \ref{f:iso}. The first thing to note is that there is a general trend that qualitatively agrees very well with the expected behaviour of stars in the PMS phase: the TWA stars are systematically brighter than the BPMG stars at any given mass, and the BPMG stars are in turn brighter than the ABMG stars which are marginally brighter than the UMaMG stars. HD 160934 appears more consistent with a BPMG age than an ABMG. We note that HD 160934 is kinematically distinct from the BPMG, with a 0\% membership probability according to the BANYAN tools, so it is highly unlikely that it is an actual BPMG member. Another clear trend in the figure is that there is considerable consistency within the different pairs of stars; e.g. stars in pairs with similar masses also have similar luminosities. Also plotted in Fig. \ref{f:iso} is an example set of isochrones from ages from 10 Myr to 300 Myr. Above 300 Myr, there is very little further evolution expected since the main sequence is generally reached by then. In Fig. \ref{f:iso}, the BCAH15 \citep{baraffe2015} isochrones are used, but we have tested a broad set of isochrones including PARSEC \citep{bressan2012} and MIST \citep{choi2016}, which all give quite similar results. All TWA and BPMG stars match  the expected theoretical isochrones well for their respective ages, perhaps with the  exception of GJ 3305 B, which appears marginally too old for the BPMG. However, for the older ABMG and UMaMG stars in this analysis, the situation is slightly more complicated. AB Dor C fits to isochrones in the $\sim$50--100 Myr range, which is consistent with the ABMG age. AB Dor A is too massive to provide meaningful constraints, but is also consistent with such an interpretation. On the other hand, GJ 2060 A, GJ 2060 B, AB Dor Ba, AB Dor Bb, J1036 B, and J1036 C all have best-fit values below these tracks, effectively placing most of them even below the expected main sequence. The GJ 2060 pair still appears consistent with $\sim$100~Myr ages in the figure, but we note that the large error bars for those components are primarily caused by the large fractional error in the mass ratio (24\%). The fractional error in total mass is much smaller (9\%), so while it would be possible for either one of the components to be individually consistent with a 100 Myr isochrone, it would not be possible for both components simultaneously. This is shown more clearly in \citet{rodet2018}, where an isochronal analysis is performed for the GJ 2060 pair as a whole.

Many different potential explanations exist for why a data point may deviate from the isochronal expectation, but in this population analysis several of them  appear unlikely. For the case of J1036, an incorrectly estimated distance could cause it to deviate from the theoretical expectation, but this explanation would not work for AB Dor, where component C is consistent with the expectations, while Ba and Bb are not. An incorrect age estimation is another general factor of importance, but it cannot explain the most deviant cases here, since they are below the main sequence and would not fit  an existing theoretical isochrone of any age. A high mass relative to the observed brightness could also in principle be caused by unresolved additional components in the systems. However, for GJ 2060 A, the B component is already clearly impacting the radial velocity of its stellar lines, and any closer companion of similar mass would impose an even greater velocity amplitude except in the case of pathological inclinations, so it would be very difficult to hide additional companions to GJ 2060 A. Furthermore, both the AB Dor Ba/Bb and J1036 BC pairs have internally similar luminosities and masses, so if we hide a companion to one component, we must in principle also hide a similar companion to the other in order to conserve the consistency between the observed components. This seems to be an overly complicated solution. The isochrones assume roughly solar composition, so a significantly different metallicity could cause a deviation from these isochrones, but this is not expected in the solar neighbourhood, and for example does not explain  the case of J1036 BC, where a range of metallicities were tested in the models \citep{calissendorff2017}. One possible explanation is missing physics in the isochronal models, which might impose incorrect slopes or offsets in the isochrones. For example, it is certainly possibly to imagine an ad hoc line in Fig. \ref{f:iso} that would function as a satisfactory isochrone for all of the ABMG members. Another hypothesis, as argued in e.g. \citet{azulay2017a}, is that the high stellar activity for low-mass stars alters the radii and effective temperatures of the components from the theoretical expectations. In this scenario, however, it is somewhat surprising that the deviation from the isochrones is largest among the oldest stars in our sample, even though such stars are statistically the least active.

\section{Summary and conclusions}
\label{s:summary}

We have presented an orbital analysis  of the AB Dor Ba/Bb pair based on several new epochs of astrometry from NACO-SAM and ZIMPOL, which have been added to existing literature astrometry. This allows us to set significantly tighter constraints on all orbital parameters than has previously been possible. Much like several other YMG binaries analysed in the literature, we find that the stars are more massive than  would have been predicted based on their brightnesses from isochronal analysis. We further address this peculiar trend by uniformly plotting all known low-mass YMG binaries with known individual stellar masses in mass--luminosity space. The targets in TWA and BPMG appear to be reasonably internally consistent, and can be matched by sensible theoretical isochrones for their expected ages. For ABMG and older targets, there is a similar internal consistency, but they generally do not match   their theoretical isochrones well. 

Using mass and bolometric luminosity as the fundamental parameters for isochronal analysis is a comparatively straightforward and model-free approach, and thus avoids many uncertainties that can otherwise arise, such as uncertainty in the SpT-$T_{\rm eff}$ relation if $T_{\rm eff}$ is used as an observable. The prospects for this kind of analysis is improving rapidly with time, for several reasons. Firstly, \textit{Gaia} will provide precise distances to all M dwarfs in the solar neighbourhood. This will improve error bars in both dynamical mass and bolometric luminosity. Furthermore, the \textit{Gaia} astrometry will improve YMG membership assessments, and probably identify new members suitable for isochronal analysis. In addition, many M-dwarf binaries have now been monitored for well over a decade with high-resolution imaging \citep{janson2014b}, which means that orbital constraints will soon become available even for moderately wide binaries of $\sim$40~yr orbital periods. The fact that many of these young binaries are also significantly radio emitting \citep{azulay2017a} is another factor that brings additional potential since it allows for closer pairs to be resolved with VLBI than is possible at visible/infrared wavelengths and can provide absolute astrometry allowing for mass ratio determinations and independent parallax estimations. Furthermore, VLBI imaging is relatively insensitive to day-night cycles, and could thus potentially reach orbital epochs that are unattainable to visible/infrared imaging during the times of the year when the target is on the dayside of Earth. This is particularly relevant for a case such as AB Dor Ba/Bb, with a period only 5 days shorter than an Earth year, such that certain parts of its orbital phase are hidden from visible light imaging continuously for over a decade.

\begin{acknowledgements}
M.J. gratefully acknowledges funding from the Knut and Alice Wallenberg Foundation. The authors thank J. Guirado and R. Azulay for useful discussion, and the anonymous referee for useful comments. We also thank the ESO staff for their efficient support. This study made use of the CDS services SIMBAD and VizieR, and the SAO/NASA ADS service.
\end{acknowledgements}

\end{document}